\documentclass{emulateapj}

\begin{document}

\title{Taking Advantage of the COS Time-Tag Capability: Observations of
  Pulsating Hot DQ White Dwarfs and Discovery of a New One}

\author{P. Dufour\altaffilmark{1}, S. B\'eland\altaffilmark{2},
  G. Fontaine\altaffilmark{1}, P. Chayer\altaffilmark{3}, and 
  P. Bergeron\altaffilmark{1} }

\altaffiltext{1}{D\'{e}partement de Physique, Universit\'{e} de
  Montr\'{e}al, C.P. 6128, Succ. Centre-Ville, Montr\'{e}al,
  QC H3C 3J7, Canada; dufourpa,fontaine,bergeron@astro.umontreal.ca}
 
\altaffiltext{2}{Center for Astrophysics and Space Astronomy, 389 UCB,
  University of Colorado, Boulder, CO 80309, USA;sbeland@colorado.edu}

\altaffiltext{3}{Space Telescope Science Institute, 3700 San Martin
  Drive, Baltimore, MD 21218, USA; chayer@stsci.edu}

\begin{abstract}

  We present an analysis of the ultraviolet light curves of five Hot
  DQ white dwarfs recently observed with the Cosmic Origins
  Spectrograph (COS) on board the Hubble Space Telescope (HST). These
  light curves were constructed by extracting the time-tag information
  from the FUV and NUV spectroscopic data. Single-color light curves
  were thus produced in 60 s time bins. The Fourier analysis of these
  data successfully recovers the main pulsation modes of the three
  stars previously known to be variable from ground-based
  observations. We also report the discovery of pulsations in another
  object, SDSS J1153+0056, making it only the fifth member of the new
  class of variable Hot DQ stars, and the first pulsating white dwarf
  to be discovered from space-based observations. The relatively high
  amplitudes of the modes observed in the FUV -- 2 to 4 times that
  observed in the optical -- as well as the high fraction of stars
  variable in our sample suggest that most, if not all, Hot DQ white
  dwarfs might be pulsating at some level when observed at high enough
  sensitivity. Our results also underline the vast potential of the
  time-tag capability of the HST/COS combination.

\end{abstract}


\section{ASTROPHYSICAL CONTEXT}

White dwarfs stars are traditionally found to have surface
compositions made primarily of hydrogen or helium. However, a new
family has recently been uncovered, the so-called Hot DQ white dwarfs,
which have surface compositions dominated by carbon with little or no
trace of hydrogen and helium \citep{dufournat,dufour08}. These stars are
exceedingly rare, and only 14 Hot DQ white dwarfs have been found
among the sample of spectroscopically-identified white dwarfs which now
comprises more than 12,000 objects \citep{dufour10}. It is believed that
the Hot DQ's could be the progenies of stars similar to the unique hot
PG~1159 star H1504+65, and could represent a new evolutionary sequence
that follows the asymptotic giant branch \citep{dufournat}. It
is also possible that they evolved from stars that were initially
massive enough to ignite carbon and form oxygen/magnesium/neon cores
with a carbon/oxygen envelope. 

Given the range of effective temperature in which Hot DQ white dwarfs
are found (18,000-24,000 K, according to Dufour et al. 2008), and
proceeding in analogy with the presence of instability strips along the 
white dwarf cooling sequences (the ZZ Ceti strip centered around 12,000
K for the H-dominated atmosphere white dwarfs, and the V777 Her strip
centered around 25,000 K for the He-dominated atmosphere white dwarfs),
\citet{montgomery} carried out a search for variability among them, 
which led to the discovery of the first pulsating Hot DQ white dwarf, SDSS 
J1426+5752. The hypothesis that the luminosity variations seen in that
star are caused by pulsational instabilities associated with low-order
and low-degree gravity-mode oscillations (as in the other known types of
pulsating white dwarfs) is backed by the exploratory nonadiabatic
calculations carried out by Fontaine, Brassard , \& Dufour (2008).  

Since then, three new members have been added to this new class of
variable white dwarf stars \citep{barlow,dunlap}, thus opening up the
exciting possibility of probing the internal structure and testing the
proposed formation scenarios of Hot DQ's through asteroseismological means.

In order to exploit fully the asteroseismological potential and better
understand the origin and evolution of Hot DQ white dwarfs, accurate
determinations of the atmospheric parameters and physical properties
(effective temperature, surface gravity, mass, surface composition) are
first necessary. Given the unusual atmospheric compositions in
particular, it is necessary to improve upon the current estimates of
these parameters. This requires better model atmospheres than currently
available, incorporating the best physics and opacities available,
notably of several carbon ions. This also requires better optical
spectroscopic observations than those available through the Sloan
Digital Sky Survey archives. Efforts on both of these fronts are
currently being made and will be presented in due time.

In addition, at the effective temperatures where Hot DQ white dwarfs
are found, most of the flux is emitted in the ultraviolet portion of
the electromagnetic spectrum. According to model atmosphere
calculations presented in \citet{dufour08}, this region is particularly
affected by the presence of numerous strong features of ionized
carbon. These absorption features have a huge impact on the energy
distribution and thermodynamic structure of the atmosphere because of
flux redistribution. Hence, it is of utmost importance to be confident
that the models correctly reproduce the UV portion of the spectrum.

In order to verify this explicitly, we requested and were awarded 33
orbits of HST time to observe Hot DQ white dwarfs with the Cosmic Origins
Spectrograph (COS). Beyond the spectroscopic capability of COS, this
instrument has also the capacity to observe in time-tag mode, making it
an ideal instrument to measure time variabilities of astronomical
objects in the ultraviolet. In the following, we thus focus on the
temporal aspect of our observations (the spectroscopic analysis will be
presented elsewhere), and we discuss the characteristics and properties
of the light curves that we extracted from our recent HST/COS
observations of five Hot DQ white dwarfs.

\section{OBSERVATIONS}\label{observation}

Our sample consists of five of the nine Hot DQ white dwarfs first analyzed
in some detail by \citet{dufour08} on the basis of optical spectroscopy. 
These objects were chosen in order to cover a representative variety of
observed Hot DQ properties, such as the presence of traces of helium,
hydrogen, magnetic fields, and variability. At the time of the
Cycle 17 call for proposals (March 2008), only SDSS~J1426+5752 was
known to exhibit pulsations and, as such, it was included in our sample
despite its faintness (g = 19.16). As a consequence, to reach the
desired minimal signal-to-noise ratio of 20 in our spectroscopic
observations, this object received the longest temporal coverage in
our sample, a fortunate situation for time variability analysis (see
below). Note also that, since the call for proposals, two of our
scheduled targets have been found to show luminosity variations from
ground-based observations, namely, SDSS~J2200$-$0741 \citep{barlow} and
SDSS~J1337$-$0026 \citep{dunlap}.

Because most of the carbon lines that dominate the spectra are very
broad, low resolution observations with the FUV G140L and NUV G230L
gratings were carried out. In order to cover as much as possible of
the spectral energy distribution, three different nominal wavelength
settings were used for the NUV observations, while one setting was
used for the FUV region. This resulted in an almost complete coverage
between $\sim$1200 and 3000 \AA~ for each star. The light curves were
extracted from the COS FUV and NUV observations taken in time-tag
mode. In this mode, each photon has a time stamp for its detection
with a resolution of 0.032 s. Before processing the FUV data, thermal
and geometric corrections were
applied\footnote{www.stsci.edu/hst/cos/documents/handbooks/current/toc.html}
\citep{beland06,McPhate}. Additionally, a pulse height filtering was
applied to the FUV data to remove low and high gain events that are
not related to actual photons. For the data used in this paper, only
the very low and very high pulse height events were removed, keeping
all the photons with a pulse height between 4 and 30 (the full range
is between 0 and 31). For these observations, less than 0.7$\%$ of the
data was filtered out.  For both the FUV and NUV data, a dead-time
correction was also applied to compensate for the loss of counts as
the count rate increases. Most the the count rates were in the order
of 50 counts/second for a deadtime correction of around 0.04 $\%$. In
the case of the NUV data, a flat field correction was also
applied. There is currently no flat field for the FUV detector. The
default extraction boxes for the spectra and the background areas were
used with a bin size of 60 s, a compromise to get a good enough S/N on
each data point and, at the same time, to sample adequately the
pulsation cycles. Because of the much higher background noise seen
with the NUV detector, smaller extraction regions for the spectra were
chosen in this case in an attempt to minimize the noise contribution
in the resulting count rates. Also, the regions where the coronal
lines appear were removed from the extracted regions.

It turned out that all of the NUV light curves that we extracted were
too noisy to be useful, except for those associated with our brightest
target, SDSS J2200$-$0741 at g = 17.70. Indeed, we could not see obvious
luminosity variations in the NUV light curves following a cursory look, and
this was confirmed quantitatively by a Fourier analysis that failed to
reveal periodicities with significant amplitudes above the (high) noise
level in the light curves of each of the three pulsators known a priori.
In contrast, and despite the fact that the count rates were
lower in the FUV bandpass than in the NUV range, all of our FUV light
curves proved themselves useful. In particular, obvious brightness
variations are seen in the FUV light curves of the known pulsators. 

\begin{figure}[h]
\plotone{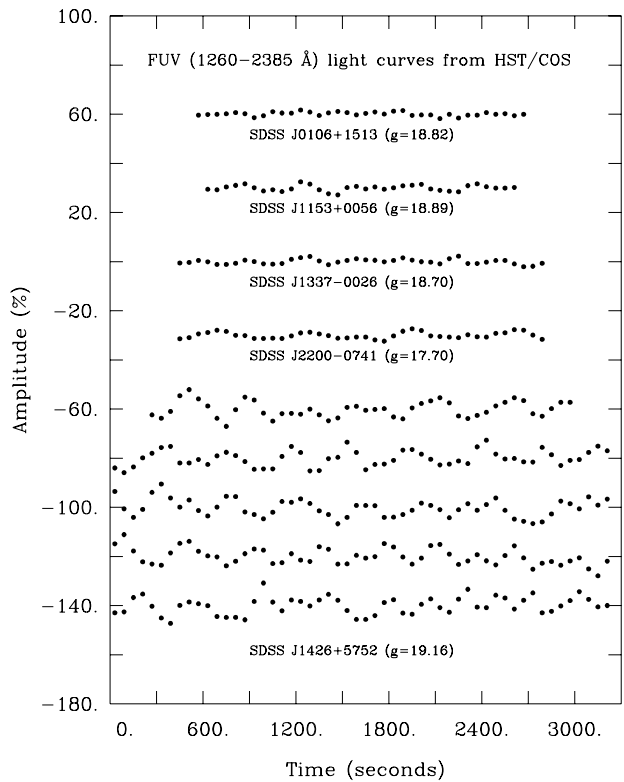}
\caption{HST/COS light curves for the five observed Hot DQ white dwarfs
  from FUV G140L data (60 s bins).}
\end{figure}

\begin{deluxetable}{lccccccc}
  \tabletypesize{\scriptsize} \tablecolumns{11} \tablewidth{0pt}
  \tablecaption{Journal of Observations}
  \tablehead{Target & Date & Start Time & Exp. Time \\ & &(UT)&(s)}
 \startdata
 J0106+1513\footnote{FUV 1260-2385\AA} & 2009-11-25  & 17:53:28 & 2164.192\\
 J1153+0056$^a$ & 2010-05-12  & 09:41:17 & 2087.200\\
 J1337$-$0026$^a$ & 2010-05-23  & 22:14:19 & 2455.168\\
 J1426+5752$^a$ & 2009-12-20  & 15:51:49 & 15871.520\\
 J2200$-$0741\footnote{NUV 2104-2419\AA} & 2010-06-16  & 18:49:43 & 2579.168\\
 J2200$-$0741\footnote{NUV 1671-2072\AA} & 2010-06-16  & 20:25:36 & 2579.168\\
 J2200$-$0741$^a$ & 2010-06-18  & 21:55:14 & 2515.168\\
 \enddata
\end{deluxetable}

In the end, we retained the light curves whose characteristics are
summarized in Table 1. For each target, there is a FUV light curve
segment available (five segments in the case of the faint object SDSS
J1426+5752), and, in addition, there are two NUV light curve segments
available for SDSS J2200$-$0741. Figure 1 compares the FUV light
curves for the five target stars. Luminosity variations are very
clearly present in the three known pulsators, SDSS J1337$-$0026, SDSS
J1426+5752, and SDSS J2200$-$0741, as well as in SDSS J1153+0056, a
star that was not known to vary before. In contrast, photometric
activity seems to be at a minimum in SDSS J0106+1513, another star
whose variability status was unknown prior to this study.

\section{DISCUSSION}\label{results}

We find it quite remarkable, especially in view of the relative
faintness of our targets, that the HST/COS combination could pick up
clear luminosity variations such as those illustrated in Figure 1. In
order to verify the validity of our light curve extraction procedure,
we first sought to recover published pulsation modes for each of the
three known variable Hot DQ stars in our sample.

\begin{deluxetable*}{lccccccc}
  \tabletypesize{\scriptsize} \tablecolumns{11} \tablewidth{0pt}
  \tablecaption{Periodicities Detected in the HST/COS Light Curves of
    Hot DQ White Dwarfs} \tablehead{Star & Period & Amplitude & S/N &
    Optical Period & Optical Amplitude & S/N & Reference \\
    &(s)&(\%)&&(s)&(\%)} \startdata
  J1153+0056 & 374.4$\pm$6.2 & 1.39$\pm$0.23&4.8& \ldots  & \ldots & \ldots &this work \\
  & 159.1$\pm$1.5 & 1.06$\pm$0.23&3.7& \ldots  & \ldots & \ldots & \\
  J1337$-$0026 & 326.6$\pm$5.3 & 1.14$\pm$0.25&3.6& 331-341  & 0.23-0.40 & 4.7 & \citet{dunlap} \\
  J1426+5752\footnote{Analysis done with a single FUV light curve (3256 s)} & 410.4$\pm$6.1 & 2.99$\pm$0.64&4.4& 417.7069$\pm$0.0008  & 1.630$\pm$0.048& 27.2&\citet{green} \\
  & 210.5$\pm$2.2 & 2.16$\pm$0.64&3.2& 208.8534$\pm$0.0007  & 0.487$\pm$0.048& 8.1 & \\
  J1426+5752\footnote{Analysis done with the 5 FUV light curves (15872 s)} & 417.75$\pm$0.40 & 3.16$\pm$0.33&7.5& 417.7069$\pm$0.0008  & 1.630$\pm$0.048& 27.2&\citet{green} \\
  & 209.02$\pm$0.27 & 1.20$\pm$0.33&2.9& 208.8534$\pm$0.0007  & 0.487$\pm$0.048& 8.1 & \\
  J2200$-$0741\footnote{Analysis done with the single FUV light curve
  available (2515 s)} & 628.0$\pm$12.2 & 1.40$\pm$0.21&5.7& 654.397$\pm$0.056  & 0.800$\pm$0.036& 17.8 &\citet{dufour09} \\
  & 330.5$\pm$5.5  & 0.86$\pm$0.21&3.5& 327.218$\pm$0.017  & 0.655$\pm$0.036& 14.6 & \\
  J2200$-$0741\footnote{Analysis done with 3 light curves (FUV and NUV
  combined; 7673 s)} & 653.17$\pm$0.16 & 1.12$\pm$0.14&6.2& 654.397$\pm$0.056  & 0.800$\pm$0.036& 17.8 &\citet{dufour09} \\
  & 326.90$\pm$0.04  & 1.02$\pm$0.14&5.7& 327.218$\pm$0.017  & 0.655$\pm$0.036& 14.6 & \\
 \enddata
\end{deluxetable*}
\vskip 1cm

Because of its faintness, SDSS J1426+5752 has the best temporal
coverage (15872 s) of all targets. We show the Fourier Power Spectral
Density of the entire FUV data set (5 light curve segments) in the
upper part of Figure 2. Clearly, the light curve is dominated by a
single oscillation, and there is also a contribution from its first
harmonic. Using standard procedures, we extracted two oscillations
from the light curve, and the characteristics of these oscillations
are summarized in the middle of Table 2. We thus find a dominant
pulsation mode with a period of 417.75$\pm$0.40 s and an amplitude of
3.16$\pm$0.33\% of the mean brightness of the star in the FUV
bandpass, a result that formally corresponds to a 7.5-$\sigma$
detection. The period corresponds remarkably well to the result of
Green et al. (2009) who found a dominant periodicity of
417.7069$\pm$0.0008 s in the optical light curve of SDSS
J1426+5752. The FUV amplitude of that pulsation mode is much larger
than its counterpart in the optical domain (see Table 2), but this is
expected from theoretical considerations (see below) and actually
reinforces our findings.

\begin{figure}[h]
\plotone{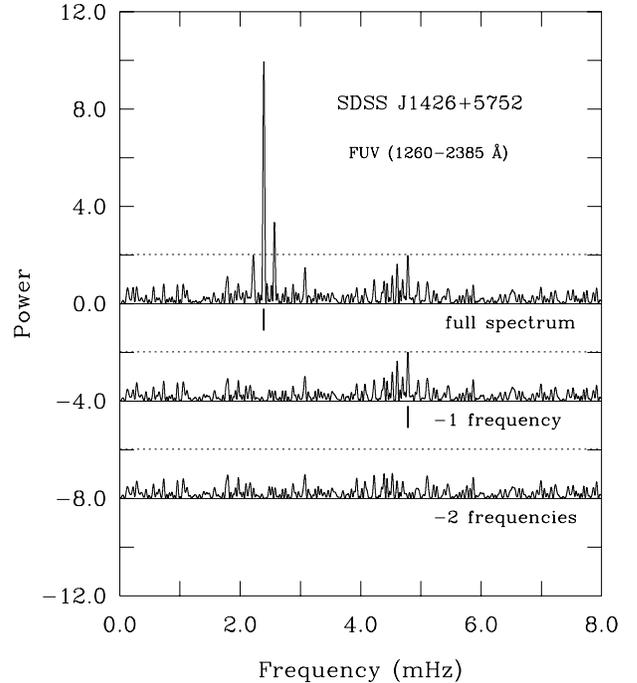}
\caption{Fourier Power Spectral Density of the entire FUV data set (5
  orbits) for SDSS~J1426+5752 in the 0$-$8 mHz range (upper
  curve). The lower transforms show the successive steps of
  prewhitening by the strongest frequency, and finally by the two
  frequencies that we isolated. The dotted horizontal lines show the
  3-$\sigma$ detection level.}
\end{figure}

Although the second oscillation picked up in our FUV light curve of
SDSS J1426+5752 is formally only a 2.9-$\sigma$ detection, it must be
considered as ``real'' given that its period of 209.02$\pm$0.27 s is
perfectly consistent with the value of 208.8534$\pm$0.0007 s found by
Green et al. (2009) in the optical bandpass. In Green et al. (2009),
as is the case here, that periodicity corresponds to the second
highest peak in the Fourier transform of the light curve of the star,
and is associated with the first harmonic of the dominant mode. Again,
the FUV amplitude of that oscillation is significantly larger than the
value of the amplitude in the optical range.

In the case of SDSS J2200$-$0741, we combined the available FUV light
curve with the two NUV light curves to obtain a total coverage of 7673
s. The results of our analysis are presented at the bottom of Table
2. We thus found two significant periodicities (653.17$\pm$0.16 s and
326.90$\pm$0.04 s) which correspond again remarkably well to the two
dominant oscillations (654.397$\pm$0.056 s and 327.218$\pm$0.017 s)
observed in the optical light curve of that star by Dufour et
al. (2009). Specifically, in both the COS spectral range and in the
optical, the light curve of SDSS J2200$-$0741 is dominated by an
oscillation and its first harmonic, the latter having an amplitude only
slightly less than its parent mode. And the amplitudes of both
periodicities are higher in the ultraviolet range than in the optical
region.

Given that the other targets, including the third known pulsator SDSS
J1337$-$0026, have only a short FUV light curve available each for
analysis, we also reported the results of another frequency extraction
exercise from a single FUV light curve for SDSS J1426+5752 in Table 2,
and similarly for SDSS J2200$-$0746. Even though the uncertainties on
the derived periods and amplitudes are substantially higher on the
basis of these shorter data sets, Table 2 indicates clearly that the
known pulsation modes are again well recovered. 

Figure 3 is a montage of the Fourier transforms of five FUV light
curves illustrated in Figure 1, one each for our five different
targets. Note that the other small peak above 3$\sigma$ near 4Mhz for
SDSS J1426+5752 in Figure 3 is a blend with the structure of the first
harmonic. After prewhitening by the dominant mode and its first
harmonic, no peaks above 3$\sigma$ remain. Similarly, no peaks above
3$\sigma$ are found in any of the 4 other light curves of SDSS
J1426+5752. We can thus be highly confident that the short light
curves are reliable and can be used to test for variability.

\begin{figure}[h]
\plotone{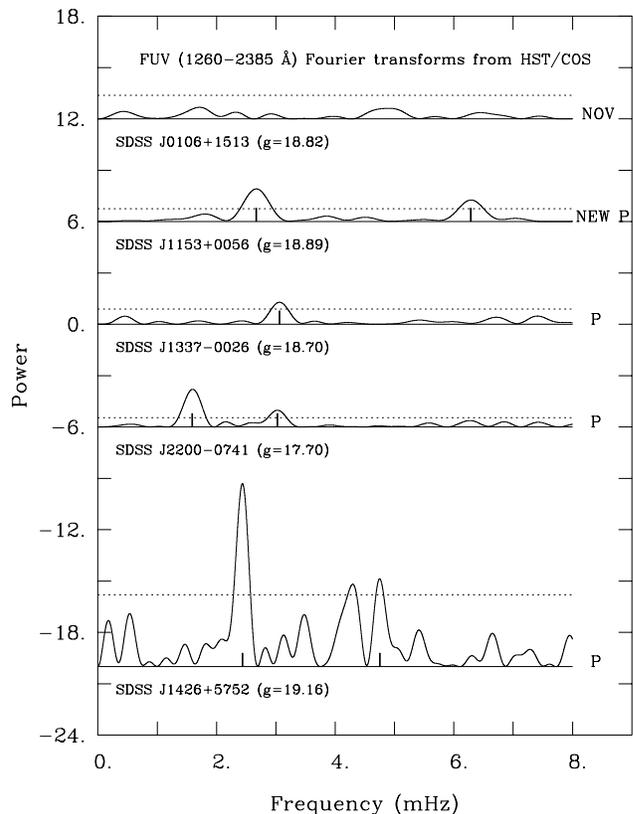}
\caption{Fourier Power Spectral Density from 5 of the FUV light curves
  shown in Fig. 1 (only the top one is used in the case of
  SDSS~J1426+5752). The dotted horizontal lines show the 3-$\sigma$
  detection level (3 times the noise level in the bandpass of interest
  after prewhitening of the detected modes). Pulsation modes are
  recovered in the three known variable stars, while oscillations are
  revealed for the first time in SDSS J1153+0056}
\end{figure}

Related to that figure, Table 2 also includes the results
of our analysis of the light curves of the three other stars in that
sample beyond SDSS J1426+5752 and SDSS J2200$-$0741. In the case of
the third known pulsating object, SDSS~J1337$-$0026, we recovered
easily (3.6-$\sigma$ detection) the main pulsation mode reported by
\citet{dunlap}. However, it is interesting to note that the high
amplitude harmonic periodicity that \citet{dunlap} found at $\sim$169
s is absent in our dataset. Had an harmonically related periodicity as
large as that reported by \citet{dunlap} been present in our dataset,
it would have been easily detected with our approach. The fact that
the amplitudes of the detected modes in all three known pulsating
targets are larger in the FUV than in the optical is a consequence of
the wavelength dependence of the limb darkening law as explained in
Fontaine \& Brassard (2008; see their Fig. 28 for instance). The
amplitude-color relationship bears the signature of the degree index
$\ell$ of a pulsation mode.

Figure 3 also reveals that SDSS J0106+1513 is not observed to vary
(NOV), at least at a detectable level (limit of 1.18$\%$ in the 0-8mHz
range), while SDSS J1153+0056 shows two distinct not harmonically
related oscillations in its light curve, consistent with what is seen
in Figure 1. The 374.4 s (159.2 s) periodicity is a 4.8-$\sigma$
(3.7-$\sigma$) detection and it has a 0.000063$\%$ (0.11\%) chance of
being due to noise according to the false alarm probability formalism
proposed by Kepler (1993). We take this as the formal proof that these
two oscillations correspond to real pulsation modes in SDSS
J1153+0056, thus making it the first pulsating white dwarf to have
been discovered from space-based observations. This underlines the
potential of the time-tag capability of the HST/COS
combination. Taking into account the rather short time bases involved
here and the fact that four out of five targets showed detectable
luminosity variations, pulsations could very well be a common
characteristic of all Hot DQ white dwarfs when observed at high enough
sensitivity.

\acknowledgements This work was supported in part by the NSERC of
Canada and FQRNT Qu\'ebec. P.D is a CRAQ postdoctoral fellow and
P.B. is a Cottrell Scholar of Research for Science
Advancement. G.F. acknowledges the contribution of the Canada Research
Chair Program. This publication makes use of data from the HST
proposal 11720 (PI: P. Dufour).

\end{document}